\documentclass{pasj01}
\usepackage{url}
\usepackage{ulem}
\usepackage{lineno}
\usepackage{comment}
\Accepted{$\langle$2023 May 31$\rangle$}

\begin{document}
\title{Nobeyama~45-m CO~$J$=1--0 Observations of Luminous Type~1 AGNs at $z\approx0.3$}
\author{Tomonari Michiyama*\altaffilmark{1,2,3,4}}%
\author{Ming-Yang Zhuang\altaffilmark{1,5,6}}%
\author{Jinyi Shangguan\altaffilmark{7}}%
\author{Hassen M. Yesuf\altaffilmark{1,8}}%
\author{Hiroyuki Kaneko\altaffilmark{9,4}}%
\author{Luis C. Ho\altaffilmark{1,5}}%
\altaffiltext{1}{Kavli Institute for Astronomy and Astrophysics, Peking University, Beijing 100871, People’s Republic of China}
\email{t.michiyama.astr@gmail.com}
\altaffiltext{2}{Faculty of Welfare and Information, Shunan University, 43-4-2 Gakuendai, Shunan, Yamaguchi, 745-8566, Japan}
\altaffiltext{3}{Department of Earth and Space Science, Osaka University 1-1 Machikaneyama, Toyonaka, Osaka 560-0043, Japan}
\altaffiltext{4}{National Astronomical Observatory of Japan, National Institutes of Natural Sciences, 2-21-1 Osawa, Mitaka, Tokyo, 181-8588}
\altaffiltext{5}{Department of Astronomy, School of Physics, Peking University, Beijing 100871, People’s Republic of China}
\altaffiltext{6}{Astronomy Department, University of Illinois Urbana-Champaign, Urbana, IL 61801, USA}
\altaffiltext{7}{Max-Planck-Institut für Extraterrestrische Physik (MPE), Giessenbachstr., D-85748 Garching, Germany}
\altaffiltext{8}{Kavli Institute for the Physics and Mathematics of the Universe (WPI), UTIAS, University of Tokyo, Kashiwa, Chiba 277-8583, Japan}
\altaffiltext{9}{Graduate School of Education, Joetsu University of Education, 1 Yamayashiki-machi, Joetsu, Niigata, 943-8512, Japan}

\maketitle

\begin{abstract}
We have performed CO~$J$=1--0 observations of ten galaxies hosting luminous ($L_{\rm bol} > 10^{46}\,{\rm erg\,s^{-1}}$) type 1 active galactic nuclei (AGNs) with the Nobeyama 45-m radio telescope. The targets are selected because they are expected to be rich in molecular gas based on their high nebular dust extinction ($A_{\rm V}$). However, no significant CO emission lines were detected in any of the targets. The upper limits of the CO~$J$=1--0 luminosities are lower than expected given the molecular gas mass inferred from the nebular $A_{\rm V}$. This inconsistency may be due to overestimated $A_{\rm V}$ values due to the lack of stellar absorption correction. 
Considering more reliable $A_{\rm V}$ values, the CO~$J$=1--0 non-detections by Nobeyama 45-m are natural. This suggests that our results do not contradict the conversion methods from $A_{\rm V}$ to molecular gas mass proposed in the literature.
This survey suggests that careful $A_{\rm V}$ measurements as well as CO observations are still needed to improve the measurements or estimates of the molecular gas content of galaxies hosting luminous AGNs.
\end{abstract}


\section{Introduction}
To understand the regulation of star formation rates and the cold gas content of galaxies, as well as the co-evolution of supermassive black holes and their galaxies,
it is important to study the feedback process associated with Active Galactic Nuclei (AGN), which can change the ionization structure and inject energy and momentum into the interstellar medium (ISM).
Studying molecular gas content in ISM could provide direct evidence for the effect of AGN feedback.
For example, fast molecular gas outflows driven by AGNs could heat and sweep out cold gas in their host galaxies, thereby inhibiting star formation and preventing the galaxies from overgrowing.
Such effects are known as negative feedback (e.g., \cite{Fabian_2012}).
On the other hand, in positive feedback, star formation activities of the host galaxy could be enhanced by star formation inside outflows driven by AGNs and/or the compression of gas due to the interaction between ISM and jet triggered by AGNs (e.g., \cite{Maiolino_2017}).

In order to understand how AGN affects their host galaxies, low-$J$ CO emissions are observed in a number of galaxies \citep{Husemann+2017MNRAS, Jarvis+2020MNRAS, Koss+2021ApJS}.
For example, \citet{Shangguan_2020a, Shangguan_2020b} investigated the molecular gas properties of a representative sample of 40 ($z\leq0.3$) quasars.
These two studies represent one of the largest and most sensitive CO surveys for low-$z$ quasars.
They found that the AGN luminosity correlates with both the CO luminosity and BH mass, suggesting that strong AGNs are gas-rich. It is necessary to increase the sample size to understand the feedback processes in gas-rich luminous AGNs,
however, measuring CO lines in large AGN samples is time-consuming.
To overcome the small sample size of CO measurements, \citet{Yesuf_2019} developed a new technique to indirectly measure molecular gas mass ($M_{\rm H_2}$) using dust extinction and metallicity in a sample of nearby star-forming galaxies (see also \cite{Yesuf2020ApJ...901...42Y,Yesuf2020ApJ...900..107Y}). 
This new method is effective because it can be applied to large surveys such as Sloan Digital Sky Survey \citep[SDSS]{York+2000AJ}.

While the methods in \citet{Yesuf_2019} have been tested on a heterogeneous sample of AGNs with CO observations \citep{ZHS_2021}, it is not clear whether this method can be applied to any AGN population, especially those with extreme dust extinction.
In this paper, we check whether this indirect method works for extremely dusty (nebular $A_V\sim4$) and luminous ($L_{\rm bol} > 10^{46}\,{\rm erg\,s^{-1}}$) type 1 AGNs from \citet{ZHS_2021} by directly observing CO~$J$=1--0 (hereafter CO~(1--0)) emission. These objects have high molecular gas mass ($>10^{10} M_{\Sol}$) predicted from their high nebular extinction. This paper is structured as follows: Section~\ref{sec:obs} and \ref{sec:res} explain the Nobeyama 45-m observations and the results, respectively. Section~\ref{sec:obs} discusses the reasons for Nobeyama 45-m non-detections. This paper assumes a cosmology with $H_0 = 70$~km~s$^{-1}$~Mpc$^{-1}$ and $\Omega_{\rm m}=0.3$.

\color{black}
\section{Nobeyama 45-m radio telescope observations}\label{sec:obs}
\subsection{Sample selection}
\begin{figure}[!htbp]
\begin{center}
\includegraphics[angle=0,scale=0.7]{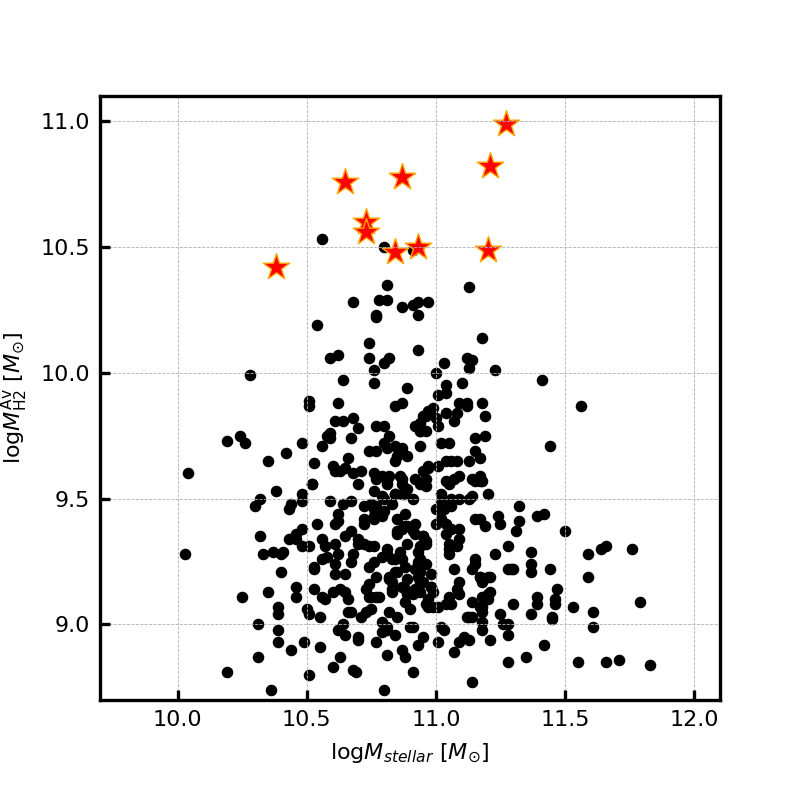}
\caption{\raggedright Predicted molecular gas mass from $A_V$ versus stellar mass for the entire $z=0.3$ type 1 AGNs sample (black dots) and objects observed by the Nobeyama 45-m (red stars).
\label{fig:sample-selection}}
\end{center} 
\end{figure}

We used the Nobeyama 45-m radio telescope to observe CO~(1--0) line of $z=0.3$ type 1 AGN sample investigated in \citet{ZHS_2021}. 
The targets are selected from 453 AGNs in \citet{ZHS_2021} based on molecular gas mass ($M_{\rm H_2}^{A_V}$) predicted by a combination of $A_V$ and metallicity using the calibration from \citet{Yesuf_2019},
\begin{equation}\label{MH2_AV}
\log M_{\rm H_2}^{A_V} (M_{\Sol}) = 0.45 A_V + 2.43 \log Z + 8.0,
\end{equation}
where $\log Z=\log({\rm O/H})-8.8$ is estimated using mass-metallicity relation from \citet{Tremonti2004} as parameterized by \citet{Kewley&Ellison2008}. 
Table~\ref{table:information} summarizes the basic properties of our targets. 
Among the targets that were observable from the Nobeyama site, we selected targets whose $M_{\rm H_2}^{A_V}$ is the largest in every observation slot assigned.
Finally, we succeeded to observe 10 AGNs (Figure~\ref{fig:sample-selection}). 
In addition, we observed PG~1700+518 (PG) to confirm that our observation setups were correct (see details in Appendix).

\begin{table*}[ht]
\centering 
\scriptsize
\begin{tabular}{cllccccccc} 
\hline 
ID & SDSS name & R.A., Decl. & $\nu_{\rm CO(1-0)}^{\rm obs}$ & z &  $M_{\rm star}$ & $\log L_{\rm bol}$ & $A_V$ & $Z$ & $M_{\rm H_2}^{A_V}$ \\
& & (J2000.0) &  (GHz) &  &  ($10^{10}~M_{\rm \Sol}$) & (erg\,s$^{-1}$) & (mag) & & ($10^{10}~M_{\rm \Sol}$) \\
(1) & (2) & (3) & (4) & (5) & (6) & (7) & (8) & (9) & (10) \\
\hline 
   Q0 & J140604.26+572956.5 & 14h06m04.3s +57d29m57s &           86.95 &         0.326 &    18.6 &         47.15 &    4.94 &             9.11 &           9.7 \\
   Q1 & J114906.09+042859.6 & 11h49m06.1s +04d28m60s &           86.51 &         0.333 &    16.2 &         46.83 &    4.56 &             9.11 &           6.5 \\
   Q2 & J012419.88+141858.5 & 01h24m19.9s +14d18m59s &           86.08 &         0.339 &     7.4 &         47.06 &    4.51 &             9.10 &           5.9 \\
   Q3 & J102700.24-010425.0 & 10h27m00.2s -01d04m25s &           85.77 &         0.344 &     4.5 &         47.42 &    4.58 &             9.08 &           5.6 \\
   Q4 & J085420.81+031144.4 & 08h54m20.8s +03d11m44s &           87.17 &         0.322 &     5.4 &         46.30 &    4.18 &             9.09 &           3.9 \\
   Q5 & J023602.08-090000.9 & 02h36m02.1s -09d00m01s &           86.30 &         0.336 &     5.4 &         45.97 &    4.09 &             9.09 &           3.6 \\
   Q6 & J161436.82+283906.0 & 16h14m36.8s +28d39m06s &           87.57 &         0.316 &     8.5 &         46.61 &    3.88 &             9.11 &           3.2 \\
  Q7* & J074613.33+332604.0 & 07h46m13.3s +33d26m04s &           87.43 &         0.318 &     6.3 &         46.43 &    3.92 &             9.10 &           3.2 \\
  Q8* & J085229.48+024713.9 & 08h52m29.5s +02d47m14s &           86.36 &         0.335 &     3.6 &         46.26 &    4.12 &             9.07 &           3.3 \\
  Q9* & J081446.27+223645.6 &  08h14m46.3s +22d36m46s &           86.93 &         0.326 &     8.1 &         46.93 &    3.85 &             9.11 &           3.1 \\
  Q10 & J204626.11+002337.7 & 20h46m26.1s +00d23m38s &           86.52 &         0.332 &     6.9 &         47.26 &    3.87 &             9.10 &           3.0 \\
  Q11 & J142441.21-000727.1 & 14h24m41.2s -00d07m27s &           87.44 &         0.318 &     2.4 &         46.47 &    4.04 &             9.04 &           2.6 \\
  Q12 & J140609.74+604500.1 &  14h06m09.7s +60d45m00s &           85.50 &         0.348 &    15.8 &         46.65 &    3.84 &             9.11 &           3.1 \\

\hline 
\end{tabular}
\caption{\raggedright 
(1) The ID number for targets. (2) SDSS name. (3) R.A., Decl.in J2000. (4) The redshifted CO~(1--0) frequency. (5) The optical redshift. (6-9) stellar mass ($M_{\Sol}$), bolometric AGN luminosity, extinction ($A_V$), and metallicity $Z=12+\log({\rm O/H})$ from \citet{Zhuang_2020}. (10) molecular gas mass ($M_{\rm H_2}^{A_V}$) estimated from nebular dust extinction in \citet{ZHS_2021}. Due to bad weather, Q7, Q8, and Q9 (marked by *) were not observed while whose $M_{\rm H_2}^{A_V}$ is larger than Q10, Q11, and Q12.} 
\label{table:information} 
\end{table*}

\subsection{Observation}
The observations were performed during the period from 2021 January to February using the Nobeyama 45-m radio telescope (project ID is CG201019). 
We used the multi-beam receiver FOREST (FOur-beam REceiver System on the 45-m Telescope; \cite{Minamidani_2016}) and the FX type correlator SAM45 (Spectral Analysis Machine for the 45-m telescope), which is equivalent to a part of the Atacama Compact Array Correlator \citep{Kuno_2011, Kamazaki_2012}.The bandwidth is 2\,GHz and the velocity resolution is 488.28\,kHz (i.e., 4096 channels).
During the observation, the pointing accuracy ($<3\arcsec$) was checked every one hour if the wind speed is $<3$~m~s$^{-1}$ and every 30 minutes if the wind speed is $>3$~m~s$^{-1}$.
This was done by observing SiO maser sources at 43 GHz or strong continuum sources (3C273). 
We observed a compact standard source IRC~+10216 during the observation to measure the main beam efficiency ($\eta_{\rm mb}$), showing that the beam efficiency is $\eta_{\rm mb}=T_{\rm a}^{*}/T_{\rm mb}=0.5-0.7$ ($T_{\rm a}^{*}$ is antenna temperature and $T_{\rm mb}$ is main beam temperature). When we investigate the upper limits of CO integrated intensity, we assume $\eta_{\rm mb}=0.5$ for conservative estimation because $\eta_{\rm mb}$ depends on pointing accuracy (see details in Appendix). For PG 1700+518, we use $\eta_{\rm mb}=0.58$ which is the value measured on 26th January.
Table~\ref{table:observation} is the summary for the observing date, sources for pointing calibration, and weather conditions.

\begin{table*}[ht]
\centering 
\footnotesize
\begin{tabular}{llcccc} 
\hline 
Date & Targets & Pointing & $T_{\rm sys}$ &  Water Vapor &  Wind  \\
(yyyy/mm/dd) & & & (K) & (hPa) & (m~s$^{-1}$) \\
    (1) & (2) & (3) & (4) & (5) & (6) \\
\hline 
2021/01/11 & J140604 (Q0) & S-Crb & 150$-$300 &$\sim 2.5$ &1$-$3\\
                    & J204626 (Q10) & O-Cet & 130$-$200 &$\sim 2.3$ &2$-$3\\
                    & J012419 (Q2) & O-Cet & 140$-$200 &$\sim 2.2$ &2$-$5\\
                    & J012419 (Q5) & O-Cet & 140$-$200 &$\sim 2.0$ &2$-$5\\
                    & J085420 (Q4) & R-Leo & 100$-$200 &$\sim 2.1$ &4$-$6\\
                    & J140604 (Q0) & S-Crb, U-Her & 140$-$300 &$\sim 2.5$ &5$-$6\\
                    & J140609 (Q12) & R-Leo & 100$-$250 &$\sim 4.3$ &5$-$7*\\
2021/01/14 & J140604 (Q0) & 3C273 & 100$-$200 &$2-3$ &1$-$2\\
2021/01/26 & J114906 (Q1) & 3C273 & 120$-$250 &$\sim 4.6$ &0$-$2\\
		      & PG~1700+518 (PG) & 3C273 & 120$-$300 & $4-5$ &1$-$2\\
2021/01/28 & J102700 (Q3) & R-Leo & 200$-$300 &$\sim 4.6$ &0$-$2\\
		      & J161436 (Q6) & U-Her & 250$-$400 &$\sim 4.8$ &2$-$4\\
		      & J142441 (Q11) & S-Crb & 160$-$240 &$\sim 4$ &3$-$5\\
\hline 
\end{tabular}
\caption{\raggedright (1) The date. (2) Target names. (3) The sources used for pointing calibration. (4) Typical systematic temperature during the observation. (5) The typical value for water vapor during the observation. (6) The typical value for wind speed during the observation. In the case of J140609 (Q12), while the wind speed is relatively high, we confirmed that pointing accuracy is $<3\arcsec$.} 
\label{table:observation} 
\end{table*}

\subsection{Data reduction}
We used Java NEWSTAR which is an astronomical data-analyzing system developed by the Nobeyama 45-m Radio Observatory. The bad scans (e.g., wavy spectra and terrible spurs) were manually flagged. Then, the baseline was fitted by linear function for each scan, and each scan for both polarization is integrated into one spectrum. 

\section{Results}\label{sec:res}
\subsection{Individual Object}
\begin{figure*}[!htbp]
\begin{center}
\includegraphics[angle=0,scale=0.36]{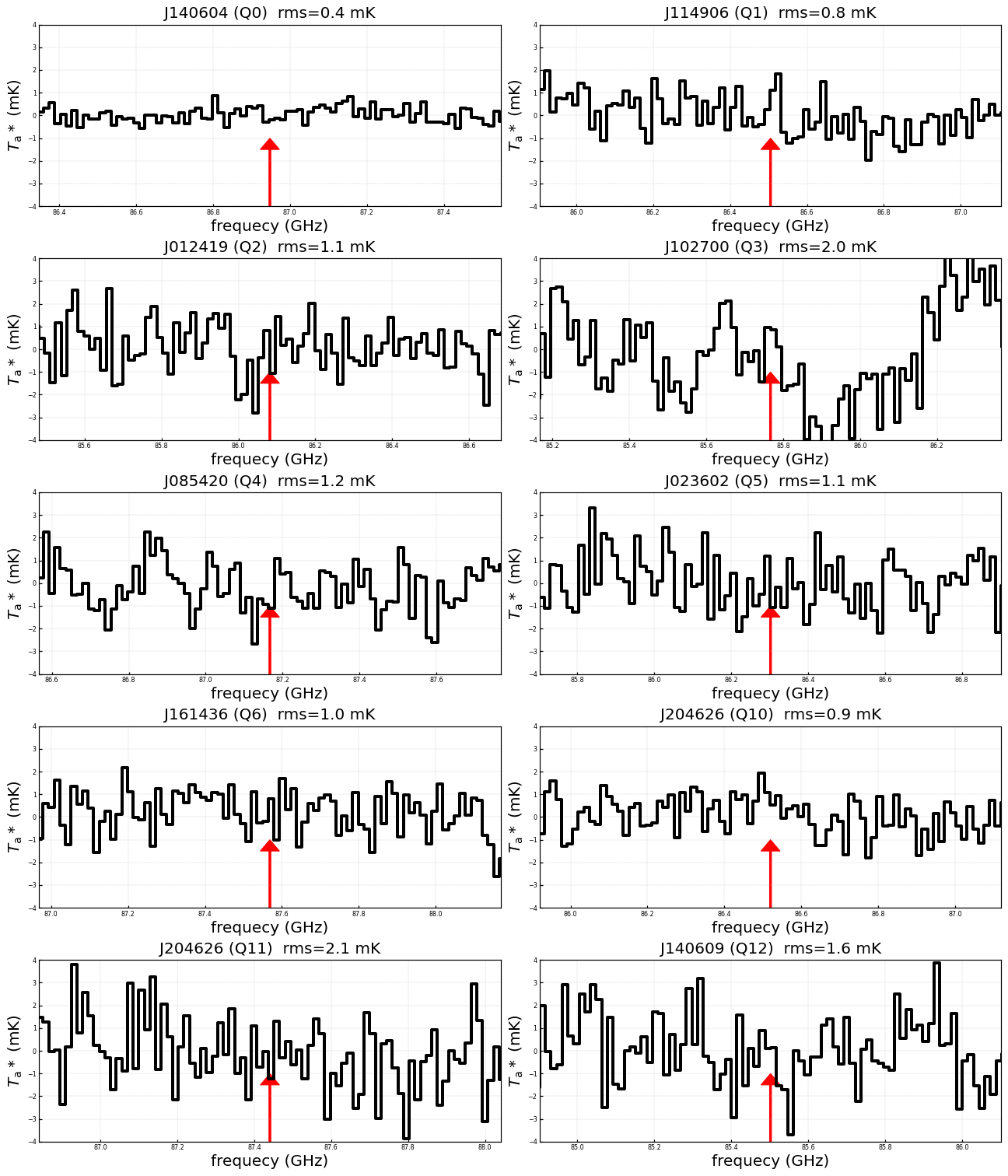}
\caption{\raggedright Each panel shows the spectra with the spectral resolution of 50~km~s$^{-1}$ for each target. Significant CO~(1--0) emissions were not confirmed in any spectrum. The red arrows indicate the redshifted CO~(1--0) frequency. 
\label{fig:spec}}
\end{center} 
\end{figure*}

After binning the channel with a velocity resolution of 50~km~s$^{-1}$, no significant CO~(1--0) emissions ($\ge 3\sigma$) were  confirmed in any spectrum (Figure~\ref{fig:spec}). Therefore, we calculate $3\sigma$ upper limit of the integrated intensity $I_{\rm CO}$ according to \citet{Hainline2004ApJ...609...61H},
\begin{equation}\label{Ico}
I_{\rm CO}~({\rm K~km~s^{-1} }) < 3 T_{\rm rms} / \eta_{\rm mb} \Delta V \sqrt{N_{\rm ch}},
\end{equation}
where 
$T_{\rm rms}$ is the rms noise in each spectra at given $T_{\rm a}^*$,  
$\eta_{\rm mb}=0.5$ is the main beam efficiency,
$\Delta V=270$~km~s$^{-1}$ is the line width assumed by the detected CO~(1--0) emission in PG~1700+618 (see Appendix),
$N_{\rm ch}=\Delta V/\delta v$ is the number of channels for the integration ($\delta v=50$\,km\,s$^{-1}$  is the velocity resolution).
Then, we calculate the $3\sigma$ upper limit of CO luminosity ($L'_{\rm CO}$) based on
\begin{equation}\label{Lco}
L'_{\rm CO}~({\rm K~km~s^{-1}~pc^{2}}) = \frac{ {\Omega_{\rm b} {I_{\rm CO}}D_{\rm L}^2}}{(1+z)^3},
\end{equation}
where $\Omega_{\rm b}$ is the beam solid angle of the main beam ($\theta{\rm mb}$=18\farcs6 at 86 GHz) and $D_{\rm L}$ is the luminosity distance (e.g., $D_{\rm L}=1730$~Mpc for J140604).
Table~\ref{table:results} presents the summary of observation information and upper limits.
In regard to J102700 (Q3), the $T_{\rm rms}$ exceeds the theoretical noise level determined from the on-source time ($T_{\rm rms}=1.1$\,mK assuming $T_{\rm sys} = 200$\,K and $t_{\rm on}=36$\,minutes). This discrepancy could be attributed to the fluctuating spectrum. The data flags by eyes may not adequately address this particular target. Nevertheless, this issue does not significantly impact the main conclusion of this paper, as we utilize upper limits for our analysis.

\begin{table*}[ht]
\centering 
\footnotesize
\begin{tabular}{llccccc} 
\hline 
Targets & $t_{\rm ON}$ & $T_{\rm rms}$ &  $I_{\rm CO}$ &  $L'_{\rm CO}$ & $M_{\rm H2}=\alpha_{\rm CO}L'_{\rm CO}$ \\
 & (min.) & (mK) & (K~km~s$^{-1}$) & ($10^{10}$~K~km~s$^{-1}$~pc$^2$) & ($10^{10}~M_{\Sol}$~) \\
    (1) & (2) & (3) & (4) & (5) & (6) \\
\hline 
J140604 (Q0) &    140 &         0.4 &          $<$0.27 &       $<$0.31 & $<$0.2, $<$1.0, $<$1.3 \\
J114906 (Q1) &     22 &         0.8 &          $<$0.55 &       $<$0.66 & $<$0.5, $<$2.0, $<$2.7 \\
J012419 (Q2) &     16 &         1.1 &          $<$0.75 &       $<$0.93 & $<$0.7, $<$2.9, $<$3.9 \\
J102700 (Q3) &     36 &         2.0 &          $<$1.37 &       $<$1.74 & $<$1.4, $<$5.4, $<$7.3 \\
J085420 (Q4) &     22 &         1.2 &          $<$0.82 &       $<$0.91 & $<$1.4, $<$5.4, $<$7.3 \\
J023602 (Q5) &     19 &         1.1 &          $<$0.75 &       $<$0.91 & $<$0.7, $<$2.8, $<$3.8 \\
J161436 (Q6) &     23 &         1.0 &          $<$0.68 &       $<$0.73 & $<$0.7, $<$2.8, $<$3.8 \\
J204626 (Q10) &     29 &         0.9 &          $<$0.62 &      $<$0.72 & $<$0.6, $<$2.3, $<$3.0 \\
J142441 (Q11) &     20 &         2.1 &          $<$1.44 &      $<$1.54 & $<$1.2, $<$4.8, $<$6.5 \\
J140609 (Q12) &     13 &         1.6 &          $<$1.09 &      $<$1.42 & $<$1.1, $<$4.4, $<$6.0 \\
\hline 
\end{tabular}
\caption{\raggedright (1) Target names. (2) On-source time after flagging the bad scan manually. (3) Rms noise level in the unit of antenna temperature (Ta$^{*}$). (4) The upper limits of integrated intensity after applying $\eta_{\rm mb}=0.5$ (i.e., the temperature corresponds to the main beam temperature $T_{\rm mb}$) calculated by equation~\ref{Ico}. (5) The CO~(1--0) luminosity calculated by equation~\ref{Lco}. (6) The upper limits of molecular gas mass assuming $\alpha_{\rm CO}=0.8, 3.1, 4.2$~$M_{\Sol}~(\rm{K~km~s^{-1}})^{-1}$, respectively.}
\label{table:results} 
\end{table*}

\begin{figure*}[!htbp]
\begin{center}
\includegraphics[angle=0,scale=0.8]{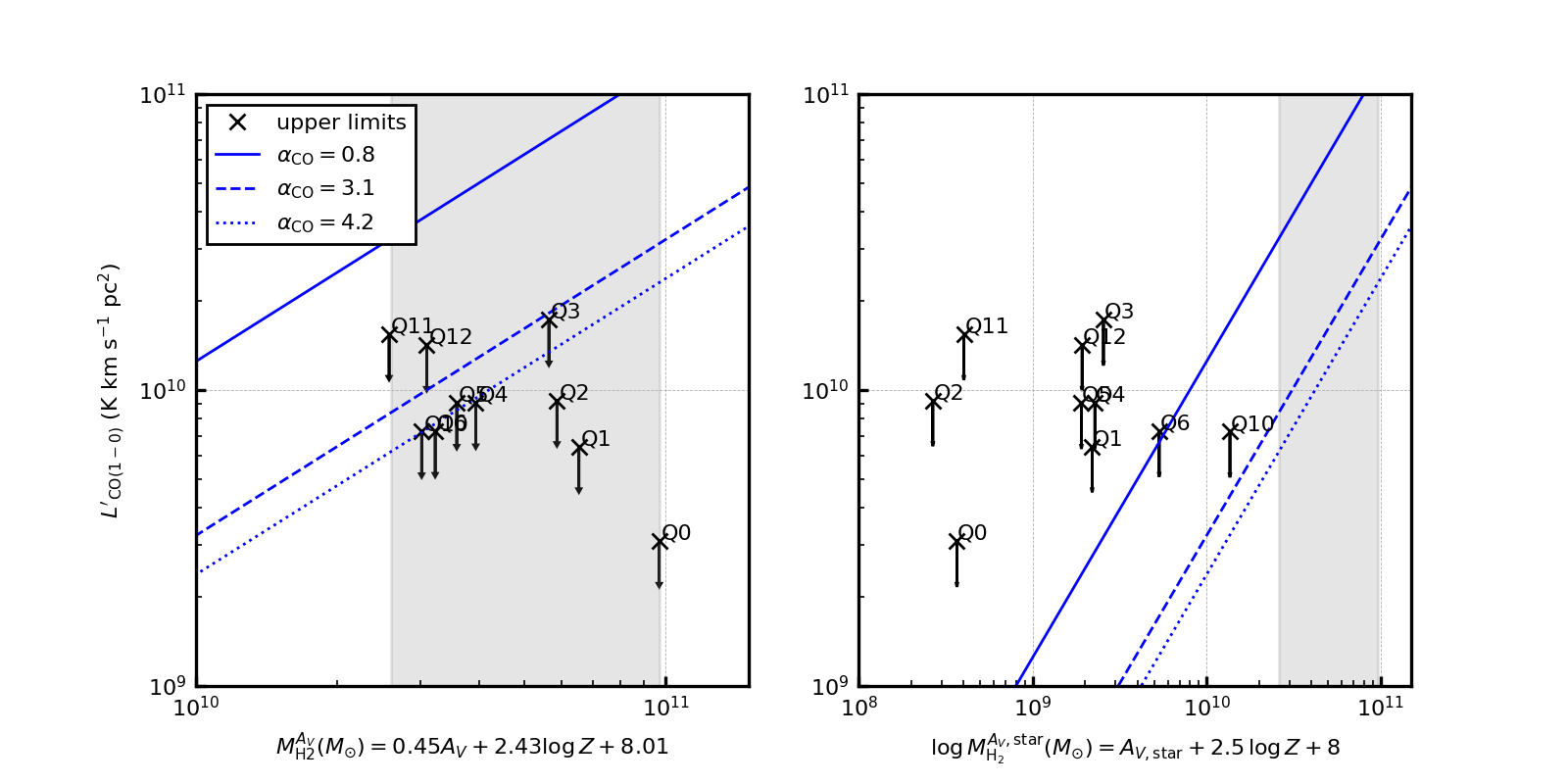}
\caption{\raggedright (left) The relation between $M_{\rm H_2}^{A_V}$ estimated in \citet{ZHS_2021} and $L'_{\rm CO}$ measured by the Nobeyama 45-m radio telescope. The blue solid, dashed, and dotted lines indicate the expected $L'_{\rm CO}$ assuming $\alpha_{\rm CO}=0.8$, 3.1 and 4.2~$M_{\Sol}~(\rm{K~km~s^{-1}})^{-1}$, respectively.
(right) Same as the left figure, but we use $\log M_{\rm H_2}^{A_V, \rm{star}} (M_{\Sol}) = \,A_{V, \rm star} + 2.5\,\log Z + 8$  (see details in section~\ref{sec:discussion}). The gray shade in both panels indicates the same mass range.
\label{fig:MH2-LCO}
}
\end{center} 
\end{figure*}

Figure~\ref{fig:MH2-LCO} (left) compares the predicted $M_{\rm H_2}^{A_V}$ with upper limits of $L'_{\rm CO}$ from our observations. The ratio between $M_{\rm H_2}$ and $L'_{\rm CO}$ is the CO-to-H$_2$ conversion factor $\alpha_{\rm CO}$ ($M_{\Sol}\ (\rm{K~km~s^{-1}})^{-1}$). 
The $\alpha_{\rm CO}\sim0.8$~$M_{\Sol}~(\rm{K~km~s^{-1}})^{-1}$ 
is often assumed for active galaxies like Ultra/Luminous Infrared Galaxies (U/LIRGs) and 
the $\alpha_{\rm CO}\sim4.2$~$M_{\Sol}~(\rm{K~km~s^{-1}})^{-1}$ 
 for normal star-forming galaxies like the Milky Way \citep{Bolatto_2013}.
Blue solid, dashed, dotted lines indicate the predicted  $L'_{\rm CO}$ for each AGNs assuming $\alpha_{\rm CO}=0.8$, 3.1, and 4.2~$M_{\Sol}~(\rm{K~km~s^{-1}})^{-1}$, respectively.
Considering typical $\alpha_{\rm CO}=3.1$ from previous AGN observations \citep{Shangguan_2020b}, the upper limits of $M_{\rm H_2}$ from CO~$J$=1--0 luminosities are lower than $M_{\rm H_2}^{A_V}$ for most targets (figure~\ref{fig:MH2-LCO} left).

\subsection{Averaged CO~(1--0) spectrum}
\begin{figure}[!htbp]
\begin{center}
\includegraphics[angle=0,scale=0.5]{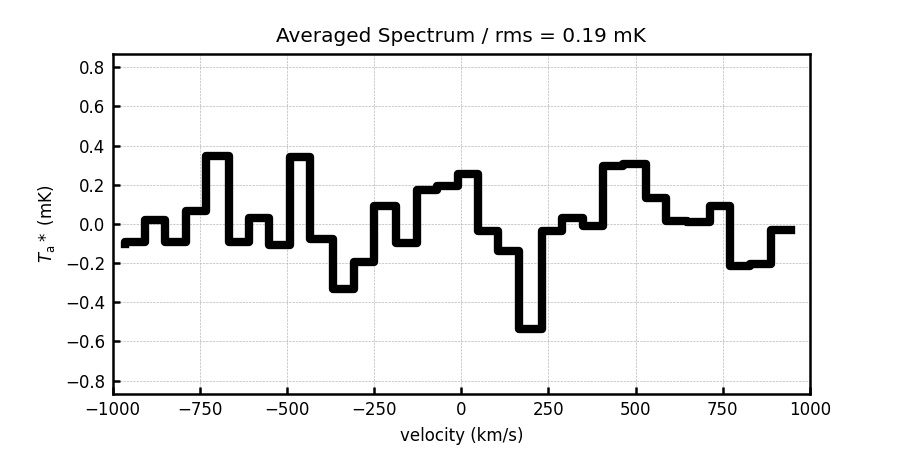}
\caption{\raggedright A noise-weighted averaged (stacked) spectrum of 10 targets at the velocity resolution of 60~km~s$^{-1}$.  \label{fig:stack}}
\end{center} 
\end{figure}

Figure~\ref{fig:stack} shows stacked (averaged by considering 1/rms weights) spectrum of ten targets.
All ten spectra were re-sampled to the frequency resolution of 5~MHz after converting the sky frequency frame into the same rest-frequency frame by using their optical redshift. Then we stacked 10 spectra without correcting the observed intensities according to the distances to galaxies.
The averaged spectrum is binned with 60~km~s$^{-1}$ velocity resolution. 
Additional baseline subtraction was done with a first-order polynomial function toward velocity averaged spectrum.
While the averaged spectrum achieved the rms noise level of 0.19~mK and three channels around the systematic velocity have positive values, the robust detection was not confirmed (no channel with the S/N$>3$).
We note that optical and CO-based redshifts of the same target could be shifted from each other. If the shifts in our targets are severe, this averaged spectrum may not imply faint CO in our targets. Therefore, we do not further discuss the non-detections in the stacked spectrum.

\section{Discussion}\label{sec:discussion}
\citet{ZHS_2021} demonstrated that equation~\ref{MH2_AV} can provide fairly good molecular gas estimates with a scatter of $\sim0.5$ dex using $\sim50$ nearby ($z<0.35$) AGNs in which CO emissions were detected in literature.
In this project, we try to confirm that this method works for very luminous AGNs with $L_{\rm bol} > 10^{46}\,{\rm erg\,s^{-1}}$ in the relatively higher-z universe ($z\sim0.3$).
However, Nobeyama~45-m CO~(1--0) non-detections show that equation~\ref{MH2_AV} may overestimate the molecular gas mass in our targets. We investigate four possible reasons for this inconsistency.

First, we investigate the possible overestimation of the nebular $A_V$ values of our targets.
The attenuation of our targets are $A_V\gtrsim 4$~mag, due to their extremely high H$\alpha$/H$\beta$ ratios of 11--15, according to \citet{2019ApJS..243...21Liu+}.
The Balmer decrement may be significantly affected by the stellar absorption, which is difficult to be measured in type~1 AGNs due to the broad emission lines.
We use equation~2 of \citet{Groves2012MNRAS.419.1402G} to estimate the stellar absorption in H$\alpha$ and H$\beta$, assuming galaxy color ($g-r\approx0.1$) for the typical starburst galaxies. Using equivalent widths (EWs) measured by \citet{2019ApJS..243...21Liu+}, the Balmer decrements become mean (85\%) H$\alpha$/H$\beta$$= 2.7\pm1.2~(3.2\pm1.2)$. If we apply $g-r\approx0.4$ for normal star-forming galaxies, H$\alpha$/H$\beta$$= 3.5\pm1.2~(4.2\pm1.2)$.  Even in the latter conservative case, $A_V\approx0.4-1.1$ in our targets. Therefore, the $A_V\approx4$ calculated in literature might be significantly overestimated.
If this correction is true, then the Nobeyama~45-m non-detections are naturally explained.
In addition, we estimated the stellar absorption ($A_{V, \rm star}$) based on UV-optical-IR spectral energy distribution (SED) fitting (Yesuf et al. in prep.), showing that the typical $A_{V, \rm star}$ value is $\approx0.6$\,mag in our Nobeyama 45-m targets. Although nebular $A_V > 4$\,mag and $A_{V, \rm star}\approx0.6$\,mag is not physically impossible, it would be unusual. If we use the $A_{V, \rm star}$ value instead, the newly predicted molecular gas mass by equation
$\log M_{\rm H_2}^{A_V, \rm{star}} (M_{\Sol}) = A_{V, \rm star} + 2.5\,\log Z + 8$
\citep{Yesuf_2019} becomes mostly $<10^{10}\,M_{\Sol}$ (figure \ref{fig:MH2-LCO} right).
Therefore, the inconsistency between the CO~(1--0) non-detection and the predicted $M_{\rm H_2}^{A_V}$ is likely due to overestimated $A_V$ values resulting from the lack of stellar absorption correction.

The second hypothes is that the equation~\ref{MH2_AV} may not work for objects with extreme nebular extinction ($A_V\gtrsim4$\,mag), which assumes $A_V\gtrsim4$\,mag is true.
The calibration sample of nearby star-forming galaxies in \citet{Yesuf_2019} only covers the range of $A_V$ from 0 -- 3 mag.
These literature AGNs have nebular $A_V=1.2^{+0.7}_{-0.6}$ mag and 
$\log \left( {M_{\rm H_2}}/{M_{\Sol}} \right)=9.3^{+0.4}_{-0.6}$, with superscript and subscript showing the difference of 84th and 16th percentiles with the median. 
The largest values are $A_V=3.4$~mag and $\log \left( {M_{\rm H_2}}/{M_{\Sol}} \right)=10.1$, respectively. These objects still are not as extreme as our targets observed here, i.e., $A_V>3.8$ in our targets.
In addition, the measured $A_V$ may not trace the bulk of the cold gas in host galaxies if such extinction is from the central ``compact" region. 
Using the high central $A_V$ in equation 1 of \citet{Yesuf_2019} leads to an overestimation of the host galaxy gas mass because the equation was derived using typical star-forming and quiescent galaxies in the xCOLDGASS survey; the AV values (within 1\farcs5 radius) of these galaxies are not extremely high.
However, to prove this hypothesis, we need CO~(1--0) observations deeper than Nobeyama 45-m to check the consistency between the nebular $A_V$ after stellar absorption correction and the $A_{V, \rm star}$ based on SED fitting.

Another possible explanation of the inconsistency between $M_{\rm H_2}$ measured by CO~(1--0) and $M_{\rm H_2}^{\rm Av}$ is the unusually large $\alpha_{\rm CO}$; i.e., $>4.2$~$M_{\Sol}$~(K~km~s$^{-1}$~pc$^2$)$^{-1}$. 
Such a large $\alpha_{\rm CO}$ can be seen in the low metallicity environment \citep{Bolatto_2013}.
However, the large $A_V$ and high metallicity estimated from the mass-metallicity relation (table~\ref{table:information}) seem to contradict the low metallicity environment scenario. 
This means that large $\alpha_{\rm CO}$ may not be a feasible reason to explain the inconsistency between $L'_{\rm CO}$ and  $M_{\rm H_2}^{A_V}$. 

Finally, we investigate the technical issues, i.e., Nobeyama 45-m beam efficiency for point sources.
We have observed PG~1700+518 for the confirmation of the telescope setup. A factor of two difference in flux is found between our measurement ($L'_{\rm CO}=(0.74\pm0.11)\times10^{10}$~K~km~s$^{-1}$~pc$^{-2}$) and that measured using IRAM-30m telescope ($L'_{\rm CO}=(1.4\pm0.2)\times10^{10}$~K~km~s$^{-1}$~pc$^{-2}$) from \citet{Evans_2009}. However, as described in Appendix~\ref{Appendix}, we had the best efforts to check the beam efficiency in each observation run.
We confirmed that correct and conservative main beam efficiency is used in our analysis. Therefore CO non-detections are not likely due to technical issues.

\section{Summary}
We observed CO~(1--0) in ten $z\approx0.3$ luminous type~1 AGNs by the Nobeyama 45-m telescope.
While a rich molecular gas reservoir is predicted from nebular dust extinction in literature, we do not detect CO~(1--0) emission in any of our targets. 
The inconsistency between prediction and observation may be due to overestimation of $A_V$ in the literature.
Our results suggest that direct observation of molecular gas using ``CO" is still necessary to 
understand the relationship between very luminous ($L_{\rm bol} > 10^{46}\,{\rm erg\,s^{-1}}$) AGN and molecular gas in the host galaxy.

\begin{ack}
T.M. and H.K. appreciate support from NAOJ ALMA Scientific Research Grant Number 2021-17A and 2020-15A, respectively.
T.M. is supported by JSPS KAKENHI grant No. JP22K14073. M.-Y.Z. and L.C.H acknowledge support from the National Science Foundation of China (11721303, 11991052, 12192220, and 12192222) and the China Manned Space Project (CMS-CSST-2021-A04 and CMS-CSST-2021-A06).
H.Y. was supported by JSPS KAKENHI Grant Number JP22K14072 and the Research Fund for International Young Scientists of NSFC (11950410492).
\end{ack}

\appendix
\section{PG 1700+518 and IRC~+10216}\label{Appendix}

\begin{figure}[!htbp]
\begin{center}
\includegraphics[angle=0,scale=0.3]{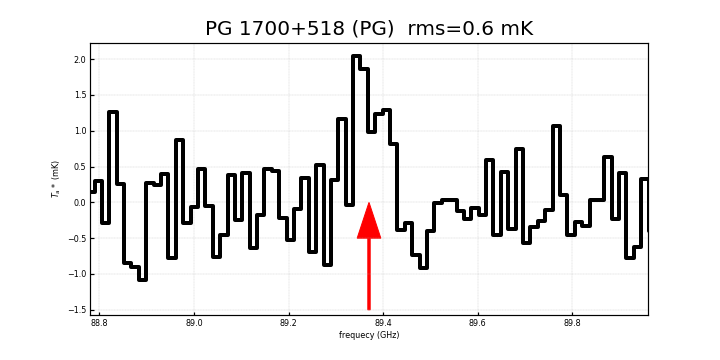}
\caption{\raggedright The CO~(1--0) detection for PG~1700+518. The red arrow indicates the frequency where CO~(1--0) line is detected by IRAM~30m. \label{fig:spec_pg}}
\end{center} 
\end{figure}

We observed PG~1700+518 to confirm that our observation setups were correct. 
In PG 1700+518, CO~(1--0) was detected by IRAM~30m telescope at 89.370~GHz (z=0.290)  with the CO~(1--0) luminosity of  $L'_{\rm CO}=(1.4\pm0.2)\times10^{10}$~K~km~s$^{-1}$~pc$^{-2}$  \citep{Evans_2009}.
Figure~\ref{fig:spec_pg} is the spectrum obtained by us during this project, showing that CO~(1--0) emission is detected at the same frequency of 89.370~GHz.
The spectrum shows that peak antenna temperature is 2~mK, the rms noise level is $T_{\rm rms}=0.6$~mK, and velocity width is FWHM=270~km~s$^{-1}$ (This FWHM is consistent with the measurements by IRAM~30m). 
The integrated intensity is $I_{\rm CO}=0.47\pm0.07$~K~km~s$^{-1}$ in the unit of antennae temperature.
Assuming $\eta_{\rm mb}=0.58$, the corresponding CO~(1--0) luminosity is $L'_{\rm CO}=(0.74\pm0.11)\times10^{10}$~K~km~s$^{-1}$~pc$^{-2}$. 
This Nobeyama 45-m measurement is $\sim50~\%$ of the value measured by IRAM~30m, which possibly suggests that the main beam efficiency is less than $\eta_{\rm mb}=0.58$.

\begin{figure}[!htbp]
\begin{center}
\includegraphics[angle=0,scale=0.5]{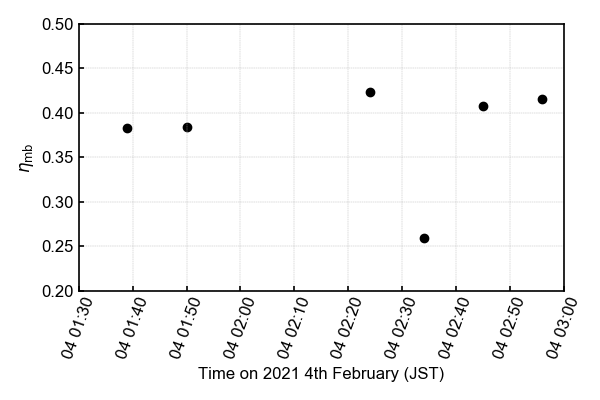}
\caption{\raggedright The time variation of $\eta_{\rm mb}$ for 90~min without pointing calibration under the strong wind speed of $\sim10$~m~s$^{-1}$ on 4th February 2021. \label{fig:eta}}
\end{center} 
\end{figure}

While $\eta_{\rm mb}$ was measured by the standard source IRC~+10216 during the observation, it significantly depends on the wind due to unstable pointing accuracy.
In order to demonstrate how stable the beam efficiency is during the observation after pointing calibration, we observed IRC~+10216 when the wind is constantly strong ($\sim10$~m~s$^{-1}$) on 4th February 2021 (We did not observe science targets due to strong wind). 
Figure~\ref{fig:eta} shows the time variation of $\eta_{\rm mb}$ for 90~min without pointing calibration. 
The typical value of $\eta_{\rm mb}=0.4$  is smaller than $\eta_{\rm mb}=0.5$ (due to bad pointing accuracy). While the efficiency was sometimes very small $\eta_{\rm mb}=0.25$ probably due to sudden strong winds, the efficiency is stable after the unexpected event.
During the observation run for science targets, we performed pointing calibration every 30 or 60 min and the wind velocity is basically $<5$~m~s$^{-1}$.
We conclude that smaller main beam efficiency is not the main reason for the inconsistency between Nobeyama 45-m telescope and IRAM 30m telescope.
We use $\eta_{\rm mb}=0.5$ when we investigate the upper limit of CO luminosity in this paper.

\bibliographystyle{aa}
\bibliography{NRO45m_AGN}

\end{document}